\documentclass[usenatbib]{mn2e}

\usepackage{graphicx}
\usepackage[space]{grffile}
\usepackage{latexsym}
\usepackage{longtable}
\usepackage{multirow,booktabs}
\usepackage{amsfonts,amsmath,amssymb}
\usepackage{natbib}
\usepackage{url}
\usepackage{hyperref}
\hypersetup{colorlinks=false,pdfborder={0 0 0}}

\newif\iflatexml\latexmlfalse
\usepackage[utf8]{inputenc}
\usepackage[english]{babel}
\usepackage[normalem]{ulem}

\title[MeqSilhouette : A mm-VLBI simulator]{MeqSilhouette : A mm-VLBI observation and signal corruption simulator}

\author[Blecher et al.]{Tariq Blecher, Roger Deane, Gianni Bernardi, Oleg Smirnov\\
~\\
\textit{Centre for Radio Astronomy Techniques and Technologies, Department of Physics and Electronics, Rhodes University,-}\\
 \textit{Grahamstown 6140, South Africa}\\
\textit{Square Kilometre Array South Africa, Pinelands 7405, Cape Town, South Africa}
}

\begin{document}

\maketitle

\label{firstpage}

\bibliographystyle{mnras}

\begin{abstract}
\newline
The Event Horizon Telescope (EHT) aims to spatially resolve the silhouette (or shadow) of the supermassive black holes in the Galactic Centre (Sgr~A$^\star$) and M87. The primary scientific objectives are to test general relativity in the strong-field regime and to probe accretion and jet-launch physics at event-horizon scales. This is made possible by the technique of Very Long Baseline Interferometry (VLBI) at (sub)millimetre wavelengths, which can achieve angular resolutions of order $\sim10~\mu$-arcsec. However, this approach suffers from unique observational challenges, including scattering in the troposphere and interstellar medium; rapidly time-variable source structure in both polarized and total intensity; as well as non-negligible antenna pointing errors. In this, the first paper in a series, we present the \textsc{MeqSilhouette} software package which is specifically designed to accurately simulate EHT observations. It includes realistic descriptions of a number of signal corruptions that can limit the ability to measure the key science parameters. This can be used to quantify calibration requirements, test parameter estimation and imaging strategies, and investigate systematic uncertainties that may be present. In doing so, a stronger link can be made between observational capabilities and theoretical predictions, with the goal of maximising scientific output from the upcoming order of magnitude increase in EHT sensitivity. 

\end{abstract}

\section{Introduction}\label{sec:intro}

The principal goal of the Event Horizon Telescope (EHT) is to spatially resolve the gravitationally-lensed photon ring (or `silhouette') of a supermassive black hole \citep{Doeleman_2010}. The two primary targets are Sgr~A$^\star$ and M87, which are expected to have gravitationally-lensed photon rings with apparent angular diameters of $\theta_{\rm pr} \sim 50$ and $\sim 20-40\ \mu$-arcsec \citep*{Falcke_2013,Broderick_2009} respectively.  The extreme angular resolution required, blurring effects due to scattering by the interstellar medium \citep[ISM; e.g.][]{Fish_2014}, and the transition to an optically thin inner accretion flow at (sub)mm-wavelengths \citep{Serabyn_1997,Falcke_1998}, necessitates that the EHT be comprised of antennas separated by thousands of kilometres that operate at high radio frequency ($\nu>200$~GHz).

Performing what is known as Very Long Baseline Interferometry (VLBI) at mm-wavelengths presents unique calibration challenges, including short atmospheric coherence times that are typically $\lesssim$10~s \citep{Doeleman_2009}, low calibrator source sky density, complex and variable calibrator source structure, and antenna pointing accuracies that are a non-negligible fraction of the antenna primary beam. These effects may place significant limitations on the sensitivity, image fidelity, and dynamic range that can be achieved with mm-VLBI.  Furthermore, unaccounted for systematic and/or non-Gaussian uncertainties could preclude robust, accurate Bayesian parameter estimation and model selection analyses of accretion flow \citep[e.g.][]{Broderick_2016} and gravitational physics \citep[e.g.][]{Broderick_2014, Psaltis_2016}, two of the EHT's many objectives.

Over the past decade, significant effort has been placed on advanced radio interferometric calibration and imaging algorithms for centimetre and metre-wave facilities in response to the large number of new arrays in construction or design phase, including MeerKAT, Australian Square Kilometre Array Pathfinder (ASKAP), Square Kilometre Array (SKA), Low-Frequency Array (LOFAR), and the Hydrogen Epoch of Reionization Array (HERA). A leading software package in this pursuit is \textsc{MeqTrees}\footnote{https://ska-sa.github.io/meqtrees/} \citep*{Noordam_2010}, which was developed to simulate, understand and address the calibration issues to be faced with the greatly enhanced sensitivity, instantaneous bandwidth, and field-of-view of such facilities. \textsc{MeqTrees} is rooted in the Measurement Equation mathematical formalism \citep{Hamaker_1996}, which parametrises the signal path into distinct $2 \times 2$ complex  matrices called Jones matrices. This formalism and applications thereof are laid out in \citep{Smirnov_2011a,Smirnov_2011b,Smirnov_2011c} and are arbitrarily generalized to model any (linear) effect, including undesired signal corruptions that often may have subtle, yet systematic effects. \textsc{MeqTrees} has been applied to correct for direction dependent calibration errors to Karl. G. Jansky Very Large Array (VLA) and Westerbork Synthesis Radio Telescope (WSRT) observations, achieving record-breaking high dynamic range images \citep{Smirnov_2011c}. The effectiveness provided by the Measurement Equation formalism in radio interferometric calibration provides a strong motivation to explore its application to the challenging goal of imaging a supermassive black hole silhouette with mm-VLBI.

Recently, there has been an increase in the attention given to simulating EHT observations of Sgr~A$^*$  and M87 \citep{Fish_2014,Lu_2014,Bouman_2015,Lu_2016,Chael_2016}. However, these are primarily focused on image reconstruction and assume either negligible or Gaussian distributed gain errors; perfect antenna pointing accuracy; and in most cases only Gaussian convolution to simulate ISM scattering. Clearly, as the EHT array is enhanced (and possibly expanded), so too must the interferometric simulations evolve to provide ever-more physical predictions on the confidence levels with which parameters can be extracted and hence exclude theoretical models of gravity and/or accretion flows.

Given the significant, yet surmountable, observational challenges that the EHT faces, we have undertaken a project to leverage metre and cm-wavelength simulation and calibration successes and build a \textsc{MeqTrees}-based mm-VLBI-specific software package called \textsc{MeqSilhouette}. While \textsc{MeqTrees} has not yet been used in the context of mm-wavelength observations, the framework is agnostic to higher frequency implementation as long as the Measurement Equation is appropriately constructed. \textsc{MeqSilhouette} enables realistic interferometric simulations of mm-VLBI observations in order to gain deeper understanding of a wide range of signal propagation and calibration effects. In this paper we describe the simulation framework and illustrate some of its key capabilities. These include the ability to simulate tropospheric, ISM scattering, and time-variable antenna pointing error effects. As will be demonstrated in a forthcoming series of papers, this technology will enable deeper understanding of a wide range of mm-VLBI signal propagation and calibration systematics, quantify their effect on accretion flow and gravitational theoretical model selection, and hence maximise the scientific utility from EHT observations.

The paper is organized as follows: in section~\ref{sec:basic_scat} we provide an introductory discussion on scattering theory;  in section~\ref{sec:sim} we describe the implementation of the simulator and provide demonstrations of the most important modules; section~\ref{sec:discussion} summarises our results and outlines our current plan for investigations with and future implementations of \textsc{MeqSilhouette}; and finally we conclude in section~\ref{sec:conclusion}.

\begin{figure*}
\begin{center}
\includegraphics[width=1.4\columnwidth]{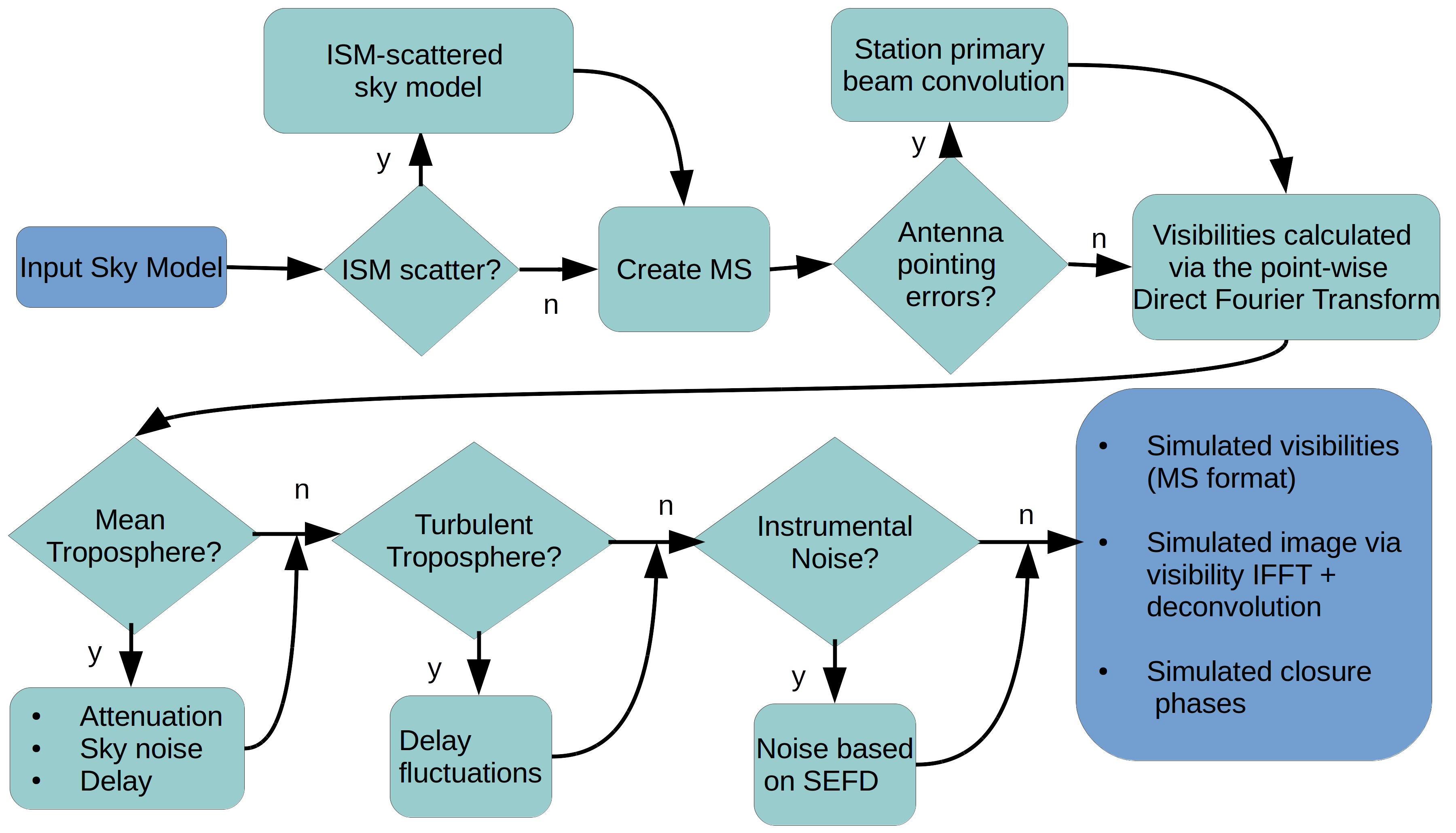}
\caption{Flow diagram showing basic sequence of the \textsc{MeqSilhouette} simulation pipeline. The sky model could include (a) a time-ordered list of {\sc fits} images or (b) parametric source model consisting of Gaussians or point sources. The details of the station information, observation strategy, tropospheric and ISM conditions are specified in a user-defined input configuration file. The pipeline is flexible, allowing any additional, arbitrary Jones matrices to be incorporated. Note that the current ISM-scattering implementation is non-linear and hence can not be incorporated into the Measurement Equation formalism. Further details in text.\label{flow}%
}
\end{center}
\end{figure*}

\section{Theoretical background}\label{sec:basic_scat}

Millimetre wavelength radiation originating at the Galactic Centre is repeatedly scattered along the signal path to the Earth-based observer. The first occurrence is due to electron plasma in the ISM \citep[e.g.][]{Bower_2006,Gwinn_2014}, while the second is due to poorly-mixed water vapour in the Earth's troposphere \citep*[e.g.][]{Carilli_1999, Lay_1997}. It is essential that the effects of the scattering phenomena are understood for accurate calibration and robust inference of the intrinsic source properties.  To this end, simulation modules approximating scattering in both media are implemented in \textsc{MeqSilhouette}. As an introduction to the separate descriptions of each, we review a simple scattering model.

An electro-magnetic wave is scattered when it passes through a medium with refractive index inhomogeneities. Following \citet{Narayan_1992}, this effect can be modeled as a thin screen, located between source and observer planes and orientated perpendicular to the line-of-sight. The screen, indexed by coordinate vector $\mathbf{x}$, adds a stochastic phase $\phi(\mathbf{x})$ to the incoming wave at each point on the screen, yielding a corrugated, outgoing wavefront. We define the Fresnel scale as  $r_{\rm F} = \sqrt{\lambda D_{\rm os}/2\pi}$, where $D_{\rm os}$ is the observer-scatterer distance, or the distance where the geometrical path difference $\frac{2\pi}{\lambda} (D_{\rm os} - \sqrt{D_{\rm os}^2 + r_{\rm F}^2}) =\frac{1}{2}$~rad.

To determine the resultant electric field at a point in the plane of the observer, indexed by coordinate vector $\mathbf{X}$, one has to take into account all possible ray paths from the screen to $\mathbf{X}$. To illustrate the model, a calculation of the scalar electric field generated by a point source, $\psi(\mathbf{X})$ yields the Fresnel-Kirchoff integral \citep*{BORN_1980}
\begin{equation}\label{Fresnel- Kirchoff}
\psi(\mathbf{X}) = C \int_{\rm screen} \exp\left[i\phi(\mathbf{x}) + i \frac{(\mathbf{x}-\mathbf{X})^2}{2 r_{\rm F}}\right]\mathbf{dx},
\end{equation}
where C is a numerical constant.

The statistical properties of $\phi(\mathbf{x})$ can be described by a power spectrum or equivalently the phase structure function,
\begin{equation}\label{eq:D_phi}
D_\phi (\mathbf{x},\mathbf{x'}) = \langle \left[ \phi(\mathbf{x} +\mathbf{x'}) - \phi(\mathbf{x})\right]^2 \rangle,
\end{equation}
where $\mathbf{x}$ and $\mathbf{x'} $ represent two points on the screen and $\langle .. \rangle$ denotes the ensemble average. 

 $D_\phi$ can be reasonably approximated by a power law dependence on the absolute distance $r$ between points on the screen 
\begin{equation}
D_\phi (r) =  (r/r_0)^\beta,\qquad r^2 = (\mathbf{x} - \mathbf{x'})^2
\label{kolmogorov}
\end{equation}
where $r_{\rm 0}$ is the phase coherence length scale defined such that $D_\phi(r_{\rm 0}) = 1$~rad. 

Kolmogorov turbulence, which describes how kinetic energy injected at an outer length scale $r_{\rm out}$ cascades to increasingly smaller scales until finally dissipated at an inner length scale $r_{\rm in}$, predicts $\beta=5/3$ in the domain ${r_{\rm in}\ll r \ll r_{\rm out}}$. This scaling has been demonstrated to be a reasonable approximation for the ISM over scales $r \sim 10^2$~km to $>1$~AU \citep*{Armstrong_1995}, and also for the troposphere with $r < \Delta h$, where $\Delta h$ is the thickness of the turbulent layer \citep{Coulman_1985,carilli_1997}.

The two length scales, $r_{\rm F}$ and $r_{\rm 0}$, define the nature of the scattering which is split into the strong and weak regimes. In \emph{weak scattering}, $ r_{\rm 0} \gg r_{\rm F}$ and hence by equation~(\ref{kolmogorov}), $D_{\phi}(r_{\rm F}) \ll 1$. This implies that most of the radiative power measured on a point $\mathbf{X}$ will originate from a screen area $A_{\rm weak} \approx \pi r_{\rm F}^2$. Whereas in the regime of \emph{strong scattering}, $ r_{\rm 0} \ll r_{\rm F}$ yielding  $D_{\phi}(r_{\rm F}) \gg 1$. This  results in coherent signal propagation onto the point $\mathbf{X}$ from multiple disconnected zones each of area $A_{\rm strong} \approx \pi r_{\rm 0}^2$ \citep{Narayan_1992}. Scattering at millimetre wavelengths in the troposphere and the ISM in the direction of the Galactic Centre fall into the regimes of weak and strong scattering respectively.

To evolve the screen in time, one typically assumes a frozen screen i.e. that the velocity of the individual turbulent eddies is dominated by the bulk motion of scattering medium \citep[e.g.][]{Lay_1997}. This allows us to treat the screen as frozen but advected over the observer by a constant motion. Hence, time variability can be easily incorporated by the relative motion between source, scattering screen and observer.

\section{Central components and layout of the simulator}\label{sec:sim}

\textsc{MeqSilhouette} is an observation and signal corruption simulator written in \textsc{Python} and {\sc MeqTrees} using the  {\sc measurement set}\footnote{https://casa.nrao.edu/Memos/229.html}  data format. A flow diagram of the simulator is shown in Fig.~\ref{flow}. Input to the simulator is a sky model and configuration file. The former is typically a time-ordered list of {\sc fits} images, where each image represents the source total intensity\footnote{Later versions of {\sc MeqSilhouette} will enable the full Stokes cubes as input.} over a time interval $\Delta t_{\rm src} = t_{\rm obs}/N_{\rm src}$, where $t_{\rm obs}$ is the observation length and $N_{\rm src}$ is the number of source images. The configuration file specifies all parameters needed by the pipeline to determine the particular observation configuration (array, frequency, bandwidth, start time, etc.) and which signal corruption implementation should be employed. The visibilities are calculated through evaluation of the Fourier Transform at each UVW coordinate in the dataset, the time and frequency resolution of which is specified by the user. The primary outputs of the pipeline are an interferometric dataset in {\sc measurement set} format along with the closure phases and uncertainties and a dirty and/or deconvolved image (or spectral cube if desired). The modular structure of the pipeline allows for multiple imaging and deconvolution algorithms to be employed.  The rest of this section is devoted to describing the implementation of each signal corruption module.

\subsection{Interstellar medium}

Scattering in the ISM at millimetre wavelengths towards the Galactic Centre falls into the strong scattering regime, which can be further subdivided into \emph{snapshot}, \emph{average} and \emph{ensemble-average} regimes \citep*{Narayan_1989,Goodman_1989}. Following from section~\ref{sec:basic_scat}, patches on the scattering screen with linear size $\sim r_0$ will emit electromagnetic waves into single-slit diffraction cones of angular size $\theta_{\rm scatt} \sim \lambda /r_0$. For a point source, an observer will be illuminated by many patches spanning  $\theta_{\rm scatt}$ with projected size on the screen equal to the {\emph refractive scale},
\begin{equation}
r_{\rm ref}\ =\ \theta_{\rm scatt} D_{\rm os}\ =\ r_{\rm F}^2/r_0.
\end{equation}

The diffraction cones from each small region will interfere, resulting in a multi-slit \emph{diffractive scintillation} pattern. A single realisation of this pattern falls in the\emph{ snapshot} regime. An extended source $\theta_{\rm src} \gg r_0/R$ will average over many realisations and quench the diffractive scintillation. In the \emph{average} regime, although diffractive scintillation has been averaged over, there still exists scintillation over scales comparable to the size of the scattered image of a point source $\sim r_{\rm ref}$, termed \emph{refractive scintillation}. This scintillation acts to focus/defocus the ensemble of coherent patches of linear size $\sim r_0$. This weak, large-scale scintillation is more difficult to average over, requiring multi-epoch observations over weeks to months in order to allow the scattering material to move across the source (assuming transverse ISM velocities of a few 10s of km\,s$^{-1}$). An extended source size will quench refractive fluctuations but only when $\theta_{\rm src} \gg \theta_{\rm scatt}$. In the \emph{ensemble-average regime}, all scintillation has been averaged and the scattering is equivalent to Gaussian convolution.

An algorithm which approximates scattering in the average regime, which is relevant to VLBI observations of Sgr~A$^\star$, has been implemented in the \textsc{Python}-based \textsc{Scatterbrane}\footnote{http://krosenfeld.github.io/scatterbrane} package, based on \citet*{Johnson_2015a}. This approach extends the structure function shown in equation~(\ref{kolmogorov}) to regimes where the inner and outer turbulent scales as well as the anisotropy of scattering kernel are considered. In this framework the scattered image $I_{\rm ss}$ is approximated by `reshuffling' of the source image $I_{\rm src}$ through
\begin{equation}\label{eq:scatterbrane}
I_{\rm ss}(\mathbf{x}) \approx I_{\rm src}\left(\mathbf{x} + r_{\rm F}^2 \nabla \phi(\mathbf{x})\right),
\end{equation}
where $\nabla$ is the directional derivative. Even though $\phi(\mathbf{x})$ is only coherent to $\sim r_{\rm 0}$, $\nabla \phi(\mathbf{x})$ remains spatially coherent over much larger scales, leading to the presence of refractive substructure \citep*{Johnson_2015a}.

We include the {\sc ScatterBrane} software, which has already yielded important context for mm-VLBI observations towards Sgr~A$^\star$ \citep[e.g.][]{2016arXiv160106571O}, within the {\sc MeqSilhouette} simulation framework.Our ISM module interfaces the \textsc{Scatterbrane} code within an interferometric simulation pipeline. This module enables simultaneous use of time-variable ISM scattering and time-variable intrinsic source structure within a single framework. The user is able to select a range of options relating to the time-resolution and epoch interpolation/averaging of both. By default, if the time resolution chosen to sample the source variability $\Delta t_{\rm src}$ and screen variability $\Delta t_{\rm ism}$ are unequal, we set  
\begin{itemize}
 \setlength\itemsep{1em}
\item $\Delta t_{\rm ism}=\Delta t_{\rm src}$ \qquad \qquad if \qquad  $\Delta t_{\rm src} < \Delta t_{\rm ism}$
\item $\Delta t_{\rm ism}=R(\frac{\Delta t_{\rm src}}{\Delta t_{\rm ism}})\Delta t_{\rm src}$ \ if \qquad  $\Delta t_{\rm src} > \Delta t_{\rm ism}$,
\end{itemize}
where $R$ rounds the fraction to the nearest integer.  This modification to the ISM sampling resolution avoids interpolation between different snapshots of the intrinsic source structure. Note that even though the ISM-scattering corruption is applied in the correct causal position in the signal propagation chain, equation~(\ref{eq:scatterbrane}) is non-linear and hence can not be written in the Measurement Equation formalism. 

\begin{figure*}
\begin{center}
\includegraphics[width=1.8\columnwidth]{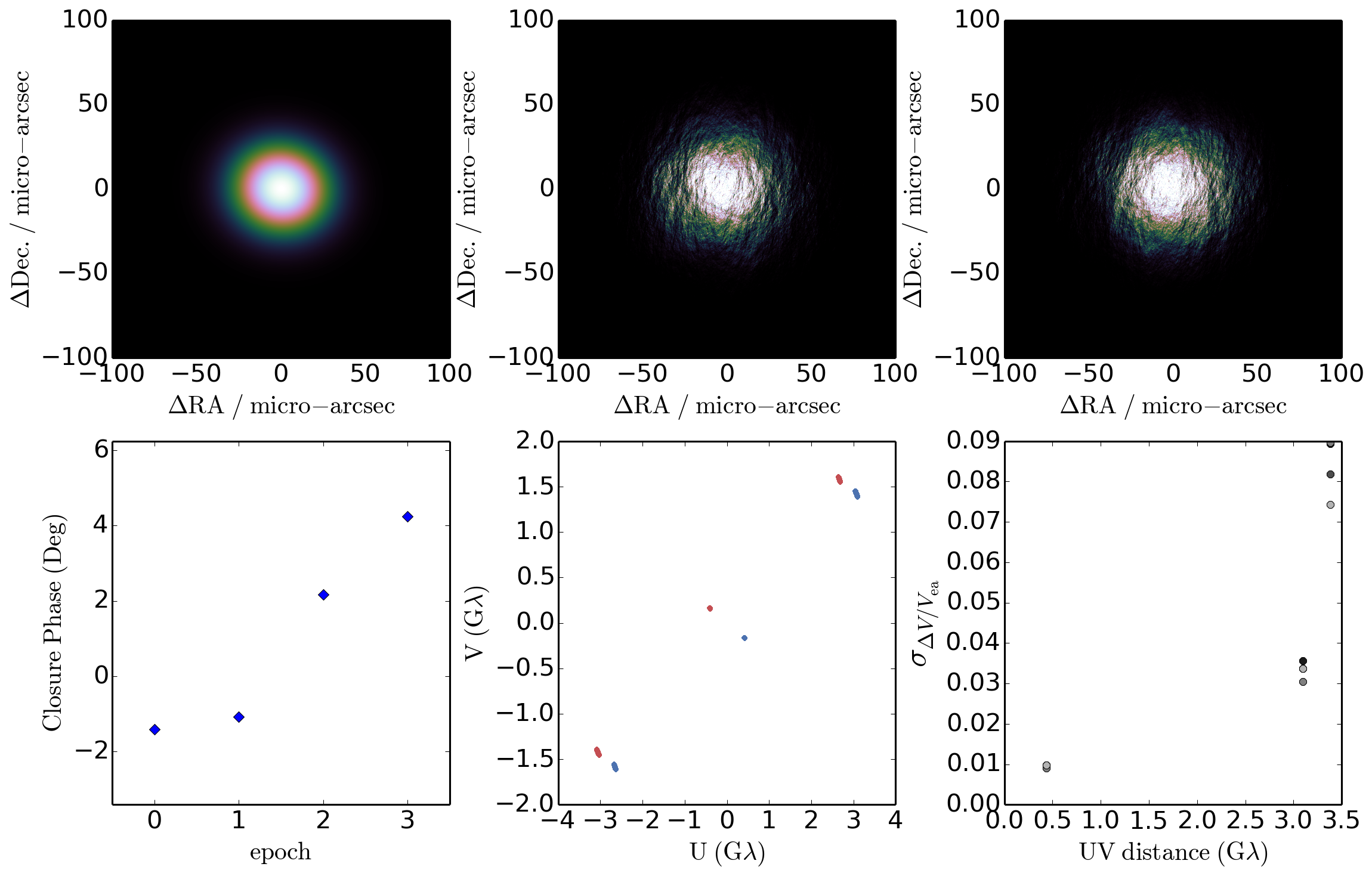}
\caption{An example simulation of ISM scattering towards Sgr~A$^{\star}$, observed with SMT-JCMT-CARMA.  The top panel, left to right, shows the original $\rm FWHM = 40$~$\mu$-arcsec Gaussian {\bf (top left)}, the simulated ISM scattered image on the first night {\bf (top middle)} and last night {\bf (top right)} of the observation, respectively.  The bottom panel, left to right,  shows the evolution of the 10 minute-averaged closure phase with epoch {\bf (bottom left)}, {\sl uv}-tracks for each night {\bf (bottom middle)} and the RMS fractional visibility amplitude differences $\sigma_{\Delta V /V_{\rm ea}}$ as a function of {\sl uv-}distance {\bf (bottom right)}. $ \Delta V= (|V_{\rm a}|-|V_{\rm ea}|)$, where $|V_{\rm a}|$ and |$|V_{\rm ea}|$ are the simulated average and ensemble average visibility amplitudes respectively. Variations from the ensemble-average flux on the shortest baselines reveal total flux modulation while flux variations on longer baselines and non-zero closure phases track the fluctuations in substructure.  Furthermore, ISM scattering simulations can constrain the variability fraction associated with the screen, enabling a more robust estimation of source variability, as demonstrated in \citet{2016arXiv160106571O}. The time-variability of the ISM is built into the {\sc MeqSilhouette} framework.\label{ISM_sequence}%
}
\end{center}
\end{figure*}

To demonstrate the implementation and provide an example of intraday ISM variability, we present the results of a simulated observation of 10 minutes duration at 14:00 UTC on four consecutive days in Fig.~\ref{ISM_sequence}. To compare to published observations, we use the three-station EHT array consisting of the Submillimeter Telescope (SMT) in Arizona, the Combined Array for Research in Millimeter-wave Astronomy (CARMA) in California and the James Clerk Maxwell Telescope (JCMT) on Mauna Kea, Hawaii. The distance to the screen is taken as $D_{\rm os}=5.8 \pm 0.3$~kpc  \citep{Bower_2014}. The relative transverse velocity between the observer and scattering screen is set to $50~\rm{km\,s}^{-1}$ to be consistent with \citet{2016arXiv160106571O}. The source is a circular Gaussian with a $\rm{FHWM}=40$~$\mu$-arcsec, approximately the angular distance that a scattering screen would travel over $\sim 4$~days. The source size has been chosen such that it is consistent with the latest estimate of the size of Sgr~A$^\star$ at $230$~GHz \citep{Fish_2011}.  Closure quantities are model dependent and calculated as specified in \citet{Rogers_1995}, where the thermal noise was added based on the system equivalent flux density (SEFD) table in \citep{Lu_2014}.

\subsection{Pointing Errors}\label{sec:pointing}

All antennas suffer pointing errors to some degree as a result of a variety of factors, including dish flexure due to gravity, wind and thermal loading, as well as imperfect drive mechanics. This corresponds to an offset primary beam, which should only translate to minor amplitude errors if the pointing error $\theta_{\rm PE}$ is significantly smaller than the primary beam (i.e. $\theta_{\rm PE} \ll \theta_{\rm PB}$). In the Measurement Equation formalism, this offset can be represented by a modified (shifted) primary beam pattern in the {\bf \it E}-Jones term 
\begin{equation}
{\bf E}_p(l,m) = {\bf E}(l_0 + \delta l_p, m_0 + \delta m_p),
\end{equation}
where $\delta l_p, \delta m_p$ correspond to the directional cosine offsets.

\begin{figure}
\begin{center}
\includegraphics[width=1.\columnwidth]{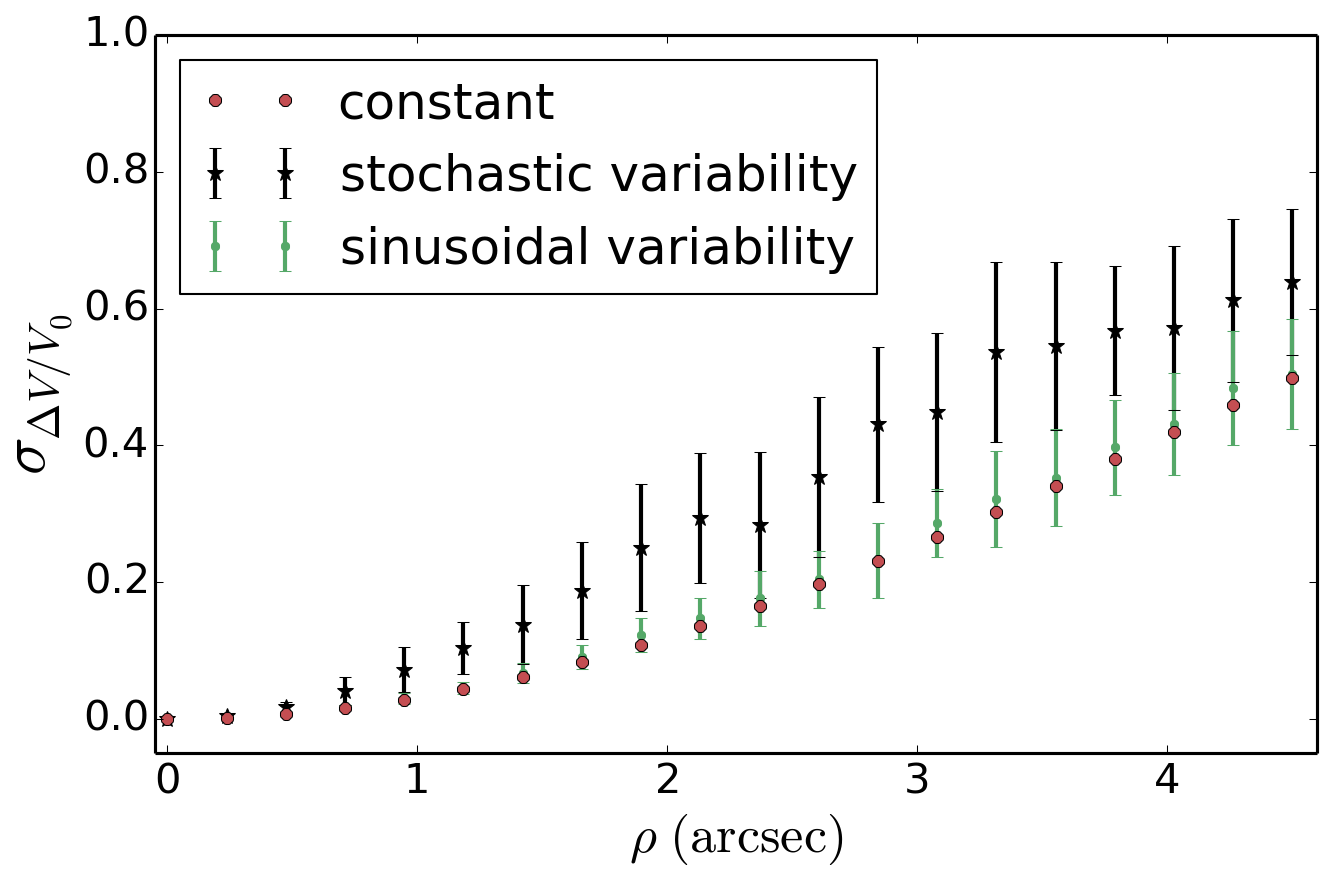}
\caption{RMS fractional amplitude error $\sigma_{\Delta V/V_0}$ induced by pointing error with the 50~m (i.e. fully illuminated) LMT antenna as a function of pointing error offset $\rho$ at $230$~GHz. We assume that these errors are degenerate or non-separable from the self-calibration/fringe-fitting model used. This simulation capability enables constraints on the magnitude of pointing-induced errors given a particular pointing calibration strategy. See text for more details. \label{fig:pointing}%
}
\end{center}
\end{figure}

We investigate the effect of pointing errors on the 50~m (i.e. fully illuminated) Large Millimeter Array (LMT) dish configured in an eight station VLBI array. The LMT has been measured to have an absolute pointing accuracy of $\sigma_{\rm abs} = 1-3$~arcsec, where smaller offsets occur when observing sources closer to zenith, and a tracking pointing accuracy $\sigma_{\rm track} < 1$~arcsec\footnote{http://www.lmtgtm.org/telescope/telescope-description/}. We investigate the observational effect of these errors through three different pointing error models which explore different instructive and plausible scenarios. The LMT has been singled out as this may well serve as a reference station for the EHT array given its sensitivity and central geographic location. The source used is a circular Gaussian of characteristic size $\Theta_{\rm src}=50$ $\mu$-arcsec, located at the phase centre. For this investigation, as long as $\Theta_{\rm src} \ll \theta_{\rm PB}$, the exact structure of the source is unimportant. We approximate the LMT beam profile using an analytic WSRT beam model \citep{Popping_2008} with a factor of two increase in the beam factor $C$ to take into account the increased dish size
\begin{equation}
E(l, m) = \cos^3(C\nu \rho),\qquad   \rho = \sqrt{\delta l_p^2 + \delta m_p^2}
\end{equation}
where $C$ is a constant, with value $C \approx 130$~GHz$^{-1}$. Note that the power beam $EE^H$ becomes $\cos^6$, resulting in a $\rm{FWHM} = 6.5 $~arcsec at 230 GHz. We make use of the RMS fractional visibility amplitude error $\sigma_{\Delta V/V_0}$, where $V_{\rm PE}$ and $V_{0}$ are the visibility amplitudes with and without pointing errors respectively, and  $\Delta V = V_{\rm PE} - V_{0}$ . In Fig.~\ref{fig:pointing}, $\sigma_{\Delta V/V_0}$ is plotted against pointing error $\rho$ over the range $0 \le \rho \le 4.5$~arcsec.

In the first case we assume a \emph{constant} pointing error. This simulation is meant to be instructive as to the typical amplitude error in the simplest possible scenario.

Also interesting to consider is a slower, continuous time-variable pointing error associated with the tracking error $\sigma_{\rm track}$. Physically, this could be attributed to changes in wind, thermal and gravitational loading which all change with telescope pointing direction and over the course of a typical few hour observation. Using the MeqTrees software package, such behaviour has been demonstrated to occur with the WSRT \citep{Smirnov_Calim_2011,Smirnov_2011c}. This is modeled as \emph{sinusoidal variability} with period sampled from a uniform distribution between 0.5 and 6 hours, and a peak amplitude $A_{\rho} = \sqrt{2} \sigma_{\rho}$ , where the factor $\sqrt{2}$ relates the peak amplitude to the RMS of a sinusoidal, zero-mean waveform.

Whilst a stationary phase centre is tracked, the pointing error should evolve slowly and smoothly, however, in mm-VLBI observations the phase centre is often shifted to another source/calibrator. This could cause the pointing error to change abruptly, with an absolute pointing error $\sim \sigma_{\rm abs}$. Source/calibrator change is scheduled every 5-10 minutes in a typical millimetre observation. An important point is that even though the EHT will be able to determine the pointing offset when observing a calibrator with well known source structure, when the antennas slew back to a source (e.g. Sgr~A$^\star$) with less certain or variable source structure, the pointing error could change significantly. This is exacerbated by the scarcity of mm-wavelength calibrators, which are often widely separated from the source. The antenna pointing error would induce scatter in the visibility amplitudes, which may be difficult to decouple from other effects e.g. intrinsic source variability and/or structure as well as time variable ISM scattering. We simulate this \emph{stochastic variability} by re-sampling the pointing error every 10 minutes from a Gaussian of characteristic width equal to the quoted pointing error. We perform 50 realisations of the simulation for each pointing offset to generate reasonable uncertainties.

In this simulation, we only consider LMT pointing errors due to its narrow primary beam and potential to be used as a reference station. However, the capability to simulate independent pointing errors for each station is available. In the case of a phased array, a pointing error simulation could be used to investigate the contribution of the pointing error to a variable phasing efficiency, which can be reasonably approximated by a scalar Jones matrix.

\subsection{Troposphere}

The coherence and intensity of millimetre wavelength electromagnetic waves are most severely deteriorated in the lowest atmospheric layer, the troposphere, which extends up to an altitude of $7-10$~km above sea level and down to a temperature $T \sim 218$~K \citep{Thompson_2001}. The troposphere is composed of primary gases $\rm N_2$ and  $\rm O_2$, trace gases (e.g. water vapour and $\rm CO_2$), as well as particulates of water droplets and dust. 
The absorption spectrum in the GHz range \citep[e.g.][]{Pardo_2001} is dominated by several transitions of $\rm H_2O$ and $\rm O_2$ as well as a pseudo-continuum opacity which increases with frequency. The pseudo-continuum opacity is due to the far wings of a multitude of pressure-broadened water vapour lines above 1~THz \citep{Carilli_1999}.

In contrast to the other atmospheric chemical components, water vapour mixes poorly and its time-variable spatial distribution induces rapid fluctuations in the measured visibility phase at short wavelengths. The water vapour column density is measured as the depth of the column when converted to the liquid phase and is referred to as the precipitable water vapour (PWV). The PWV is, via the real component of the refractive index, directly proportional to phase offset, 
\begin{equation}
\delta\phi \approx \frac{12.6\pi}{\lambda} \times w, 
\end{equation}\label{eq:phi-pwv}

\noindent where $w$ is the depth of the PWV column \citep*{Carilli_1999} and an atmospheric temperature $T=270$~K has been assumed. This relationship between phase and water vapour content has been experimentally verified \citep{hogg_1981}. At 230~GHz, the change in PWV needed to offset the phase by 1~rad is $\Delta w\approx0.03$~mm. In the mm-VLBI case, this sensitive dependence of phase coherence on atmospheric stability is aggravated by typically low antenna elevation angles, uncorrelated atmospheric variations between stations, and the sparsity of the array.

Our focus is to model three primary, interrelated observables which are the most relevant to mm-VLBI: turbulence-driven fluctuations in the visibility phase $\delta \phi$; signal attenuation due to the atmospheric opacity $\tau$; and the increase in system temperature due to atmospheric emission at a brightness temperature $T_{\rm atm}$.

Our approach is to model these observables as being separable into mean and turbulent components which are simulated independently. The mean tropospheric simulation module performs radiative transfer with a detailed model of the electromagnetic spectrum of each atmospheric constituent. The turbulent simulation module takes a scattering approach to account for the decoherence that results from power-law turbulence.

\subsubsection{Average Troposphere}
The problem of radiative transfer through a static atmosphere is well described and implemented by the Atmospheric Transmission at Microwaves (\textsc{atm}) software \citep{Pardo_2001}. \textsc{atm} has been incorporated into \textsc{MeqSilhouette} to provide a fast and sophisticated procedure to calculate average opacities, sky brightness temperatures and time delays. Here we provide a brief summary of the theory underpinning the package but refer the reader to \citet{Pardo_2001} for more detail. \textsc{atm} is commonly used in the Atacama Large Millimeter Array (ALMA) community \citep{Curtis_2009,Nikolic_2013} and has been tested with atmospheric transmission spectra taken on Mauna Kea \citep{Serabyn_1998}.

We start from the unpolarised radiative transfer equation, which is unidirectional in the absence of scattering,
\begin{equation}
\frac{dI_\nu (s) }{ds} = \epsilon_\nu(s) -\kappa_\nu(s)  I_\nu (s),
\end{equation}\label{eq:rad_trans}
where $s$ is the coordinate along the signal path through the atmosphere, $I_\nu(s)$ is the specific intensity, $\epsilon_\nu$ is the macroscopic emission coefficient and $\kappa_\nu$ is the macroscopic absorption coefficient.

The goal is to integrate this equation over the signal path which requires $\kappa_\nu$ as a function of altitude and frequency. The integration naturally yields the mean opacity and sky brightness temperature. The mean time delay is calculated from $\kappa_\nu$ using the Kramers-Kronig relations. In practice, this involves a triple sum over altitude layer, chemical species and rotational energy transition. Atmospheric temperature and pressure profiles are calculated based on several station dependent inputs, namely, ground temperature and pressure and the precipitable water vapour column depth.

A general equation to determine the absorption coefficient for a transition between a lower $l$ and upper $u$ states is given in the original paper. Here we merely point out that it should be proportional to the energy of the photon, $h\nu_{l \to u}$, the transition probability or Einstein coefficient, $ B_{l \to u}$, the line-shape, $f(\nu,\nu_{l \to u})$ and the number densities $N$ of electronic populations. Line profiles which describe pressure broadening (perturbations to the Hamiltonian due to the presence of nearby molecules) and Doppler broadening are used. The condition of detailed balance further requires that decays from the upper state are included yielding, $g_u B_{u \to l} =g_l B_{l \to u}$, where $g$ is the degeneracy of the electronic state. Putting this together we find,

\begin{equation}
\kappa(\nu) _{l \to u}  \propto  h\nu   B_{l \to u}  \left(\frac{N_l}{g_l}  -  \frac{N_u}{g_u} \right) f(\nu,\nu_{l \to u}),
\end{equation}

\noindent where the Einstein coefficients are calculated from the inner product of the initial and final states with the dipole transition operator. The number densities of the two states, $N_u$ and $N_l$ in local thermodynamic equilibrium (LTE) are simply related to the local number density and temperature via Boltzmann statistics.

\begin{figure}
\begin{center}
\includegraphics[width=1.\columnwidth]{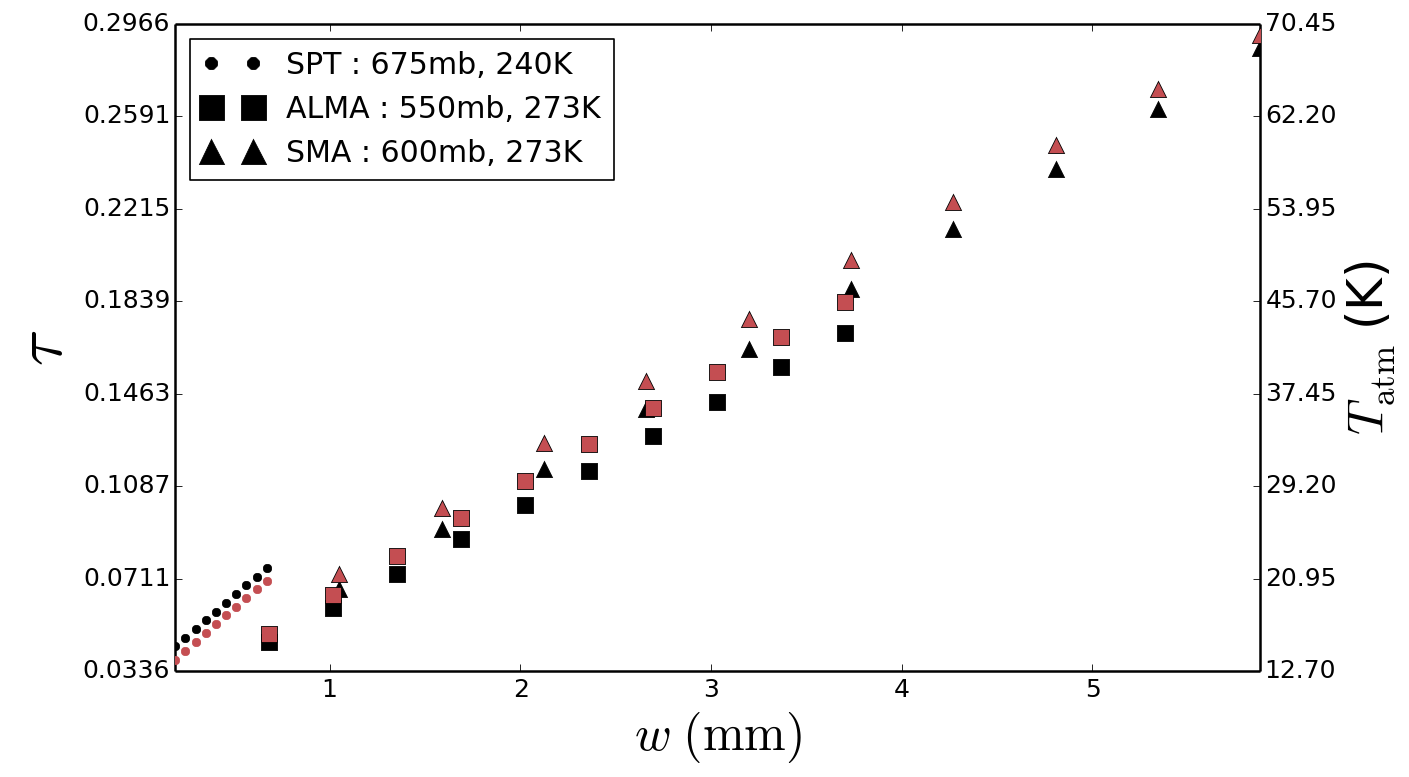}
\caption{Simulated mean opacity (black) and sky brightness temperature (red) at $\nu =230$~GHz  for three typical ground pressures and temperatures over a typical PWV range \citep{Lane_1998} which approximately represent the sites of SPT (dots), ALMA (squares) and SMA (triangles). The legend shows the estimated input ground (pressure, temperature) parameters for each site.\label{fig:mean_atm}%
}
\end{center}
\end{figure}

Typical opacities and sky brightness temperatures for ALMA, the Submillimeter Array (SMA) and the South Pole Telescope (SPT)  are shown in Fig.~\ref{fig:mean_atm}.  Note that both the opacity and brightness temperature are inversely proportional to the ground temperature and proportional to ground pressure.

\subsubsection{Turbulent troposphere}
Visibility phase instability  $\delta \phi(t)$ due to tropospheric turbulence is a fundamental limitation to producing high fidelity, science-quality maps with a mm-VLBI array \citep{Thompson_2001}. The coherence time-scale is typically too rapid ($\lesssim10$~s) for fast switching calibration, so other calibration procedures (e.g. water vapour radiometry, paired antennas, and/or self-calibration) must be performed. Self-calibration is the most commonly used but is limited by the integration time needed to obtain adequate SNR to fringe fit. Phase decoherence often leads to the use of closure quantities to perform model fitting \citep[e.g.][]{Doeleman_2001,Bower_2004, Shen_2005}.

Following from section~\ref{sec:basic_scat}, we can model the statistics of $\delta \phi(t)$ with a thin, frozen, Kolomogorov-turbulent phase screen moving at a bulk transverse velocity, $v$.  We set the height $h$ of the screen at the water vapour scale height of 2~km above ground. We will show later that the thickness $\Delta h$ of the atmospheric turbulent layer can be neglected in our implementation. At an observing wavelength of $1.3$~mm, the Fresnel scale is $r_F \approx 0.45$~m and experiments show annual variations of $r_0 \sim 50 - 500$~m above Mauna Kea \citep{Masson_1994} and $r_0 \sim 90 - 700$~m above Chajnantor \citep*{Radford_1998}, where both sites are considered to have excellent atmospheric conditions for (sub)millimetre astronomy. As $r_F < r_0$, this is an example of weak scattering.

The required field-of-view (FoV) of a global mm-VLBI array is typically FoV~$< 1$~mas or ~$\sim10~\mu$m at a height of 2~km, which is roughly 7-8 orders of magnitude smaller than the tropospheric coherence length. The tropospheric corruption can therefore be considered constant across the FoV and, from the perspective of the Measurement Equation, modeled as a diagonal Jones matrix per time and frequency interval. As VLBI baselines are much longer than the turbulent outer scale, $|\mathbf{b}| \ge 1000$~km~$\gg  r_{\rm out} \sim 10$~km, the phase variations are uncorrelated between sites and can be simulated independently. This assumption only holds for VLBI baselines and the framework needs to be extended to simulate the effects of turbulence on individual phased arrays stations (e.g. SMA) and short ($<10$~km) baselines (e.g. JCMT - SMA).

Our aim then is to produce a phase error time sequence $\left\{\delta \phi(t_i)\right\}$ for each station which is added to the visibility phase. We invoke the frozen screen assumption and write the structure function as a function of time, $D (t) =  D(r)|_{r=vt}$. The temporal structure function $D(t)$ provides an efficient route to sample the variability of the troposphere at the typical integration time of the dataset, $t_{\rm int} \sim 1$~sec. 

The temporal variance of the phase is a function of the temporal structure function, and accounting for time integration yields \citep*[see][B3]{Treuhaft_1987} 

\begin{equation}
\sigma^2_{\phi}(t_{\rm int}) = (1/t_{\rm int})^2 \int_{0}^{t_{\rm int}} (t_{\rm int}-t) D_{\phi}(t) dt.
\end{equation}

\begin{figure}
\begin{center}
\includegraphics[width=1.\columnwidth]{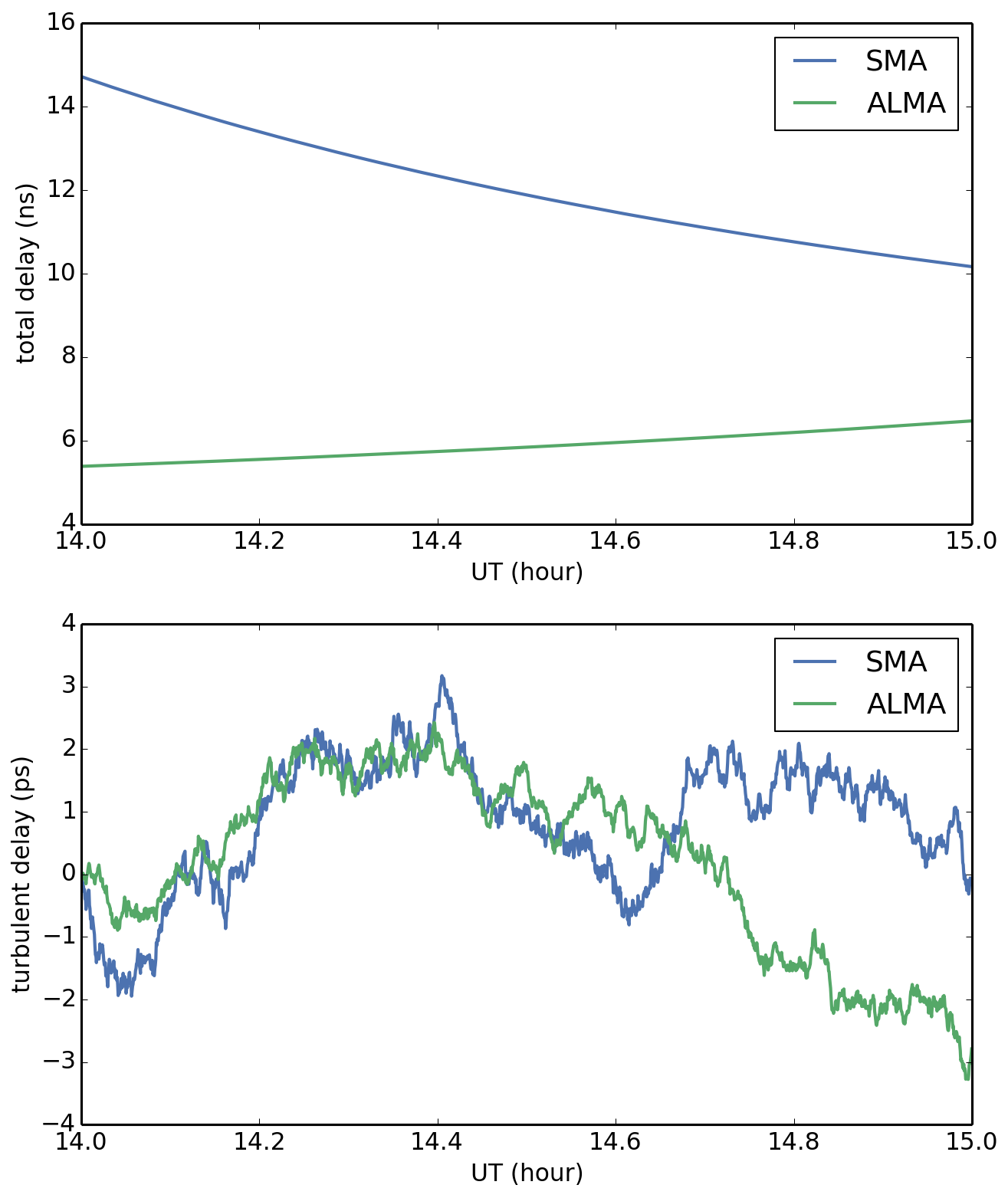}
\caption{Simulation of the total delay (top) and the turbulent atmospheric delay (bottom) for SMA (blue) and ALMA (green) sites towards Sgr~A$^\star$. Ground pressures and temperatures are the same as Fig.~\ref{fig:mean_atm}, PWV above each station is set to $w=2$~mm, and the zenith coherence time is set to $t_0=10$~s for both stations. Note that all tropospheric parameters are, however, independently set for each station. The conversion from time delay to phase at 230~GHz is $1$~rad~$\approx 0.7$~ps.\label{delay_plots}%
}
\end{center}
\end{figure}

Assuming power-law turbulence and integrating yields, 

\begin{equation}
\sigma^2_\phi (t_{\rm int})=\left[\frac{1}{\sin\theta(\beta^2 +3\beta +2)}\right]\left(\frac{t_{\rm int}}{t_0}\right)^{\beta},
\end{equation}

\noindent where $t_0 = r_0/v$ is the coherence time when observing at zenith and $1/\sin\theta$ is the approximate airmass which arises as $D_\phi \propto w$. As $r \ll \Delta h$, where $\Delta h$ is the thickness of the turbulent layer, an thin screen exponent of $\beta = 5/3$ is justified \citep*{Treuhaft_1987}. The phase error time-series takes the form of a Gaussian random walk per antenna. At mm-wavelengths, the spectrum of water vapour is non-dispersive up to a few percent \citep{Curtis_2009} and so we can assume a simple linear scaling across the bandwidth. Fig.~\ref{delay_plots} shows an example simulation of the turbulent and total delays at the SMA and ALMA sites.

Phase fluctuations $\delta\phi(t)$ can also be simulated by taking the inverse Fourier transform of the spatial phase power spectrum. However this approach is much more computationally expensive, e.g. for an observation length $t_{\rm obs}$ involving $N_{\rm ant}=8$ independent antennae with dish radii $r_{\rm dish}=15$~m, wind speed $v=10$~m\,s$^{-1}$ and pixel size equal to $r_{\rm F}$, the number of pixels $N_{\rm pix} \approx N_{\rm ant} t_{\rm obs} r_{\rm dish}^2/(v r_{\rm F}^3)  \sim 10^8$. Additionally, due to fractal nature of ideal Kolmogorov turbulence, the power spectrum becomes unbounded as the wavenumber approaches zero which makes it difficult to determine the sampling interval of the spatial power spectrum \citep{Lane_1992}.

\subsubsection{Limitations to high-fidelity image reconstruction}\label{sec:trop_errors}
A primary objective of \textsc{MeqSilhouette} is to understand and constrain systematic errors in mm-VLBI observations. In this section the tropospheric module is used to estimate the effect on image quality for various levels of calibration accuracy.

We simulate the simple scenario of a sky model that consists of a 2.4~Jy point source at the phase centre, which is the approximate EHT-measured flux density of Sgr~A$^\star$ at 230~GHz. We assume a zenith phase coherence time of $t_0=10$~s above each station (however, each stations PWV can be independently simulated). We approximate the effect of imperfect calibration by adding a small fraction of the turbulent phase noise. For this example, we do not include the mean delay component, assuming it to be perfectly corrected for during the calibration.

\begin{figure*}
\begin{center}
\includegraphics[width=1.4\columnwidth]{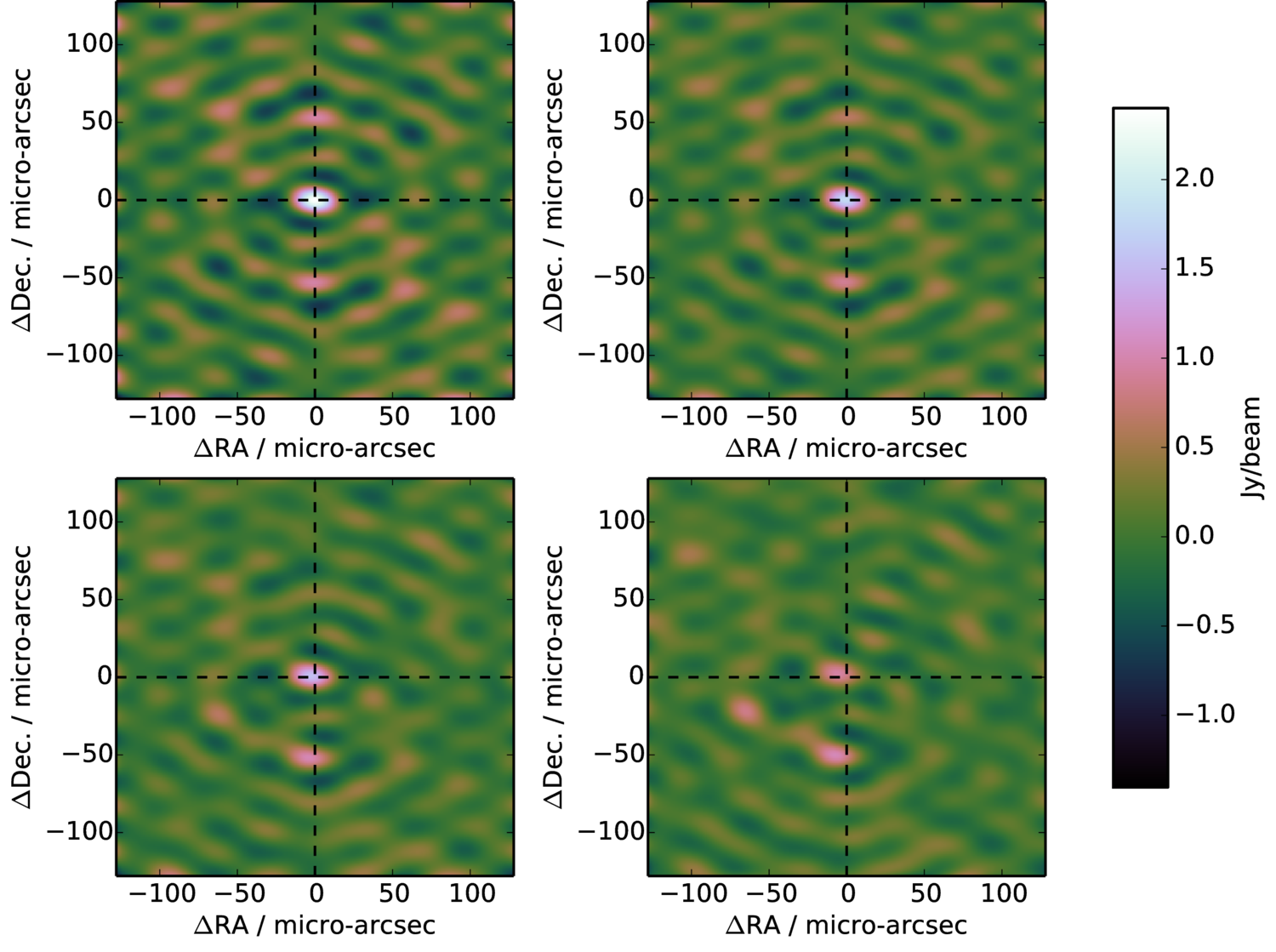}
\caption{The effect of residual troposphere phase noise on uniformly weighted images of a point source observed for 12 hours at 230~GHz with 4~GHz bandwidth with the following array: SPT, ALMA, SMA, SMT, LMT and JCMT, assuming SEFDs from \protect\citet{Lu_2014} and an elevation limit of 15$^\circ$. For simplicity the weather parameters at each station were set to: coherence time $t_{\rm 0}=10$~sec; PWV depth $w=1$~mm; ground pressure $P=600$~mb; ground temperature $T =273$~K. {\bf Top left:} Interferometric map with thermal noise only. {\bf Top right:} Atmospheric attenuation and sky noise (due to non-zero opacity) with 1\% of the turbulent phase noise added. {\bf Bottom left:} As previous, but with 3\% of turbulent phase contribution. {\bf Bottom right:} As previous, but with 6\% turbulent phase contribution. The fractional turbulent phase contributions are illustrative of the effect of fringe-fitting errors. Note the source attenuation and centroid shift that results.\label{fig:trop_images}%
}
\end{center}
\end{figure*}

\section{Discussion}\label{sec:discussion}

In section~\ref{sec:sim} we have described the layout of \textsc{MeqSilhouette} synthetic data simulation framework. A wide range of signal propagation effects can be implemented using the Measurement Equation formalism, with tropospheric scattering and antenna pointing errors given as illustrative examples. The framework is sufficiently general and flexible so that time variability in all relevant domains (source, array, ISM, troposphere) can be incorporated. The run time for a typical simulation with a realistic instrumental setup is on the order of minutes.  Implementation of polarisation effects is intended in the next version.

The ISM scattering software \textsc{ScatterBrane}, based on \citet*{Johnson_2015a}, has been incorporated into the pipeline. Fig.~\ref{ISM_sequence} provides an example of closure phase and flux variability over a 4 day period using a static source. Accurate simulation of the ISM-induced closure phase variation is essential in order to make accurate inferences regarding asymmetric, event-horizon scale structure from EHT observations \citep[e.g.][]{Fish_2016,2016arXiv160106571O}. This will become even more important as the EHT sensitivity increases by an order of magnitude in the near future. Note that if the source position is time variable as in the case of a hotspot model \citep{Doeleman_2009}, this will increase ISM variability as the relative motion between source, screen and observer is increased.

Visibility amplitude errors due to antenna pointing error has been investigated for the $50$~m  LMT dish operating at $230$~GHz. In Fig.~\ref{fig:pointing}, we show that pointing errors associated with frequent phase centre switching (stochastic variability) could introduce a RMS fractional amplitude error $\sigma_{\Delta V/V_0} \sim 0.1 - 0.4$ for an absolute pointing accuracy  $\sigma_{\rm abs} \sim 1-3$~arcsec. In contrast, tracking errors are less problematic with $\sigma_{\Delta V/V_0} \le 0.05$ for a tracking accuracy  $\sigma_{\rm track}<1$~arcsec. The case of a constant error pointing model is comparable to that of the `slow variability' case. If the gain error is non-separable from the calibration model used, it could be interpreted as intrinsic variability, substructure and/or increased noise. If unaccounted for, this effect has the potential to limit the dynamic range of mm-VLBI images. Further tests to constrain the pointing uncertainties of EHT stations will lead to more accurate interferometric simulations and hence the overall impact on black hole shadow parameter estimation. Here we demonstrate the capability to incorporate a range of plausible pointing error effects into a full simulation pipeline. For future observations at 345~GHz, these effects will be even more pronounced, given the narrower primary beam.

In section~\ref{sec:trop_errors} we explore the observational consequences of observing through a turbulent troposphere. In this simulation, we assume a simple point source model and apply increasing levels of turbulence-induced phase fluctuations before imaging using regular sampling and a two dimensional inverse fast Fourier transform. The simulated residual calibration errors result in a significant attenuation in source flux; slight offsets in the source centroid (black cross-hairs) and the presence of spurious imaging artefacts. In an upcoming paper, we perform a systematic exploration of the turbulent tropospheric effects on the accuracy of fringe-fitting algorithms and strategies, through use of an automated calibration procedure and including the added complexity of a time-variable source.

Significant progress has been made in the theoretical and numerical modeling of the inner accretion flow and jet launch regions near a supermassive black hole event horizon
\citep[e.g.][]{Zanna_2007,Etienne_2010,Dexter_2013,Moscibrodzka_2014, McKinney_2014}. As the sensitivity of the EHT stands to dramatically increase, these theoretical efforts must be complemented by advances in interferometric simulations. With \textsc{MeqSilhouette}, we now have the ability to couple these with sophisticated interferometric and signal propagation simulations.  Moreover, detailed interferometric simulations will enable us to quantify systematic effects on the black hole and/or accretion flow parameter estimation.

\section{Conclusion}\label{sec:conclusion}
In light of the science objectives of mm-VLBI observations and software advances in the broader radio interferometry community, a mm-VLBI data simulator has been developed. An important feature is that this simulation pipeline is performed using the {\sc Measurement Set} format, in line with ALMA and future VLBI data formats. The focus has been placed on simulating realistic data given an arbitrary theoretical sky model. To this end, the simulator includes signal corruptions in the ISM, troposphere and instrumentation. Examples of typical corruptions have been demonstrated, which show that each corruption can significantly affect the inferred scientific parameters. Particular focus has been placed on EHT observations, however, the pipeline is completely general with respect to observation configuration and source structure. Time variability in all domains (source, array, ISM, troposphere) is implemented.  Future versions of \textsc{MeqSilhouette} will include polarisation dependent corruptions. The creation of a close interface between sophisticated theoretical and interferometric mm-VLBI simulations will enhance the scientific opportunities possible with the EHT.

\section*{Acknowledgements}

We thank Michael Johnson and Katherine Rosenfeld for making the {\sc ScatterBrane} code publicly available, and for helpful discussions. Similarly, we thank Bojan Nikolic for \textsc{atm} support. We thank the referee for helpful comments and questions which helped to refine this work. We are grateful to Monika Mo\'{s}cibrodwska for supplying theoretical simulations we used in testing and to Ilse van Bemmel and Lindy Blackburn for useful discussions. We thank Paul Galatis for assistance with the \textsc{MeqSilhouette} logo design. The financial assistance of the South African SKA Project (SKA SA) towards this research is hereby acknowledged. Opinions expressed and conclusions arrived at are those of the author and are not necessarily to be attributed to the SKA SA\footnote{www.ska.ac.za} O.~Smirnov's research is supported by the South African Research  Chairs Initiative of the Department of Science and Technology and National Research Foundation.

\bibliography{converted_to_latex.bib%
}

\end{document}